\documentstyle[prl,floats,aps,twocolumn,epsf,graphicx]{revtex}
\begin{document}
\twocolumn[\hsize\textwidth\columnwidth\hsize\csname
@twocolumnfalse\endcsname

\title{Kerr black hole quasinormal frequencies}
\author{Shahar Hod}
\address{Department of Condensed Matter Physics, Weizmann Institute, Rehovot 76100, Israel}
\date{\today}
\maketitle

\begin{abstract}

\ \ \ Black-hole quasinormal modes (QNM) have been the subject of much 
recent attention, with the hope that these oscillation frequencies
may shed some light on the elusive theory of quantum gravity.  We
compare numerical results for the QNM spectrum of the (rotating) Kerr
black hole with an {\it exact} formula Re$\omega \to T_{BH}\ln
3+\Omega m$, which is based on Bohr's correspondence principle.  We
find a close agreement between the two. Possible implications of this
result to the area spectrum of quantum black holes are discussed.

\end{abstract}
\bigskip

]

Gravitational waves emitted by a perturbed black hole are dominated by
`quasinormal ringing', damped oscillations with a {\it discrete}
spectrum \cite{Nollert1}.  At late times, all perturbations are
radiated away in a manner reminiscent of the last pure dying tones of
a ringing bell \cite{Press,Cruz,Vish,Davis}.  The quasinormal mode
frequencies (ringing frequencies) are the characteristic `sound' of
the black hole itself, depending on its parameters (mass, charge, and
angular momentum).

The free oscillations of a black hole are governed by the well-known 
Regge-Wheeler equation \cite{RegWheel} in the case of 
a Schwarzschild black hole, and by the Teukolsky equation \cite{Teukolsky} for the 
(rotating) Kerr black hole. The black hole QNM correspond to
solutions of the wave equations with the physical boundary
conditions of purely outgoing waves at spatial infinity 
and purely ingoing waves crossing the event horizon \cite{Detwe}. Such boundary 
conditions single out {\it discrete} solutions $\omega$ (assuming a time dependence of the 
form $e^{i\omega t}$).

The ringing frequencies are located 
in the complex frequency plane characterized by Im$\omega >0$. It turns
out that for a given angular harmonic index $l$ there exist an 
infinite number of quasinormal modes, for $n=0,1,2,\dots$, characterizing oscillations with decreasing
relaxation times (increasing imaginary part) \cite{Leaver,Bach}. On 
the other hand, the real part of the frequencies approaches an asymptotic 
{\it constant} value.

The QNM frequencies, being a signature of the black-hole spacetime are of great importance from 
the astrophysical point of view. They allow a direct way of identifying the spacetime 
parameters (especially, the mass and angular momentum of the central black hole). 
This has motivated a flurry of activity with the aim of computing the spectrum of oscillations 
(see e.g., \cite{Nollert1} for a detailed review).

Recently, the quasinormal frequencies of black holes have acquired a 
different importance \cite{Dreyer,Kun,Motl,Cor,MotlNei,CarLem,KaRa} in the context of Loop Quantum 
Gravity, a viable approach to the quantization of General Relativity (see e.g., \cite{Ash,Rov} and 
references therein). 
These recent studies are motivated by an earlier work of Hod \cite{Hod1}. 
Few years ago I proposed to use 
{\it Bohr's correspondence principle} in order to determine the value of the 
fundamental area unit in a quantum theory of gravity.

 To understand the original argument it is useful to recall that 
in the early development of quantum mechanics, Bohr 
suggested a correspondence between classical and quantum properties of the
Hydrogen atom, namely that ``transition frequencies at large quantum numbers should equal
classical oscillation frequencies''. The black hole is in many senses 
the ``Hydrogen atom'' of General relativity. I therefore suggested \cite{Hod1} a similar usage of the 
discrete set of black-hole frequencies in order to shed some light on the {\it quantum} 
properties of a black hole. However, there is one important difference between the Hydrogen atom and 
a black hole: while a (classical) atom emits radiation spontaneously according to the (classical) 
laws of electrodynamics, a {\it classical} black hole does not emit radiation. This crucial 
difference hints that one should look for the highly damped black-hole free oscillations 
[let $\omega=$Re$\omega+i$Im$\omega$, then 
$\tau \equiv ($Im$\omega)^{-1}$ is the effective 
relaxation time for the black hole to return to a quiescent
state after emitting gravitational radiation. Hence, the relaxation time $\tau \to 0$ 
as Im$\omega \to \infty$, implying no radiation emission, as should be the case for a 
classical black hole].

Leaver \cite{Leaver} was the first to address to problem of computing the black hole 
highly damped ringing frequencies. Nollert \cite{Nollert2} found numerically that the asymptotic
behavior of the ringing frequencies of a Schwarzschild black hole is
given by (we normalize $G=c=2M=1$)

\begin{equation}\label{Eq1}
\omega_n=0.0874247+{i \over 2} \left(n+{1 \over 2} \right)\  ,
\end{equation}
as $n \to \infty$ \cite{Note1}. The asymptotic behavior Eq. (\ref{Eq1}) was later 
verified by Andersson \cite{Andersson} using an independent 
analysis. In Ref. \cite{Hod1} I realized that the asymptotic real part of the frequencies 
equals $\ln3 /8\pi$, and proposed a heuristic picture (based on 
thermodynamic and statistical physics arguments) trying to 
explain this fact. 
Most recently, Motl \cite{Motl} has given an analytical proof for 
this conjecture.
 
Using the relation $A=16 \pi M^2$ for the surface area of a Schwarzschild black hole, 
and $\Delta M=E=\hbar \omega$ one 
finds $\Delta A=4{\ell^2_P} \ln3$ with the emission/absorption of a quantum, where 
$\ell_P$ is the Planck length. 
Thus, we concluded that the area
spectrum of the quantum Schwarzschild black hole is given by
 
\begin{equation}\label{Eq2}
A_n=4 {\ell^2_P} \ln 3 \cdot n\ \ \ ;\ \ \ n=1,2,\ldots\ \  . 
\end{equation}

This result is remarkable from a statistical physics point of
view. It does not relay in any way on
the well known thermodynamic relation between black-hole surface area
and entropy $S_{BH}={1 \over 4}A$ \cite{Beken1}. In the spirit of Boltzmann-Einstein formula in
statistical physics, Mukhanov and Bekenstein \cite{Muk,BekMuk,Beken2}
relate $g_n \equiv exp[S_{BH}(n)]$ to the number of microstates of the
black hole that correspond to a particular external macrostate. 
In other words, $g_n$ is the
degeneracy of the $n$th area eigenvalue. The accepted thermodynamic
relation between black-hole surface area and entropy \cite{Beken1}, combined with the 
requirement that $g_n$ has to be an integer for every $n$, actually 
enforce a factor of the form $4\ln k$ 
(with $k=2,3,\dots$) in Eq. (\ref{Eq2}). We have shown that the value $k=3$ is the only one 
compatible both with the area-entropy thermodynamic relation for black hole, and with 
Bohr's correspondence principle as well.

Bekenstein \cite{Beken1,Beken2,Beken3} (see also \cite{Hod1,Hod2} and references therein) 
has given evidence for the existence of a universal (i.e., independent of the black-hole parameters: 
mass, charge, and angular momentum) area spacing for quantum black holes. 
This, combined with the universality of the black-hole 
entropy (i.e., its direct thermodynamic relation to the black-hole surface area) suggest 
that the area spectrum Eq. (\ref{Eq2}) should be valid for rotating 
black holes as well. In fact, our analysis leads to a natural 
conjecture for the asymptotic behavior of the
highly damped quasinormal frequencies of a generic (rotating) Kerr black hole. 
First, we use the first law of black-hole thermodynamics

\begin{equation}\label{Eq3}
\Delta M=T_{BH} \Delta S + \Omega \Delta J\  ,
\end{equation}
where $T_{BH}=(r_{+}-r_{-})/A$ is the Bekenstein-Hawking temperature, 
and $\Omega=4 \pi a /A$ is the angular velocity of the black-hole horizon [$r_{\pm} =M \pm (M^2-a^2)^{1/2}$ 
are the black hole (event and inner) horizons, and 
$a=J/M$ is the black hole angular momentum per unit mass]. Taking cognizance of 
Eqs. (\ref{Eq2}) and (\ref{Eq3}) [together with the relation $S={1 \over 4}A$], one finds

\begin{equation}\label{Eq4}
Re\omega \to T_{BH}\ln 3+\Omega m\  ,
\end{equation} 
where $m$ is the azimuthal eigenvalue of the oscillation. The corresponding problem of 
the Hydrogen atom in quantum mechanics hints that the formula should be valid in the 
$l=m$ case.

It should be emphasized that the asymptotic behavior of the black hole ringing
frequencies is known only for the simplest case of a Schwarzschild
black hole. Less is known about the corresponding 
QNM spectrum of the (rotating) Kerr black hole \cite{Leaver,Det,Ono}. 
This is a direct consequence of the numerical 
complexity of the problem. Onozawa \cite{Ono} computed the first nine 
frequencies of the Kerr black hole. It is of great interest to compare 
the conjectured asymptotic behavior given by Eq. (\ref{Eq4}) 
with the results of direct numerical computations.

\begin{figure}[tbh]
\centerline{\epsfxsize=9cm \epsfbox{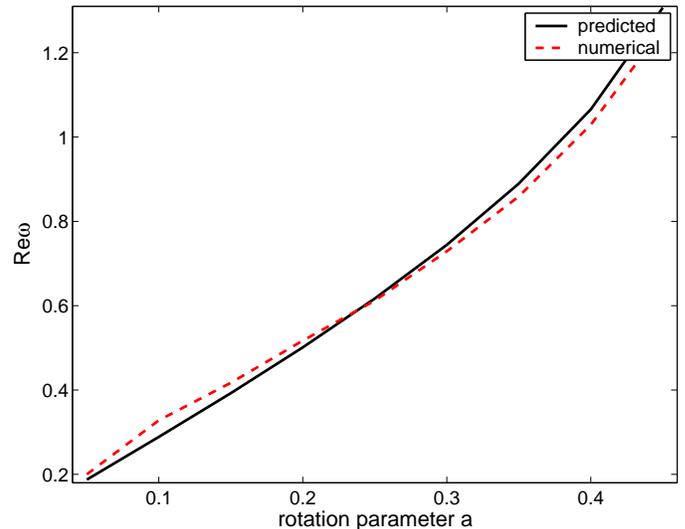}} 
\caption{Real part of the Kerr black hole QNM frequencies as a function of the black hole 
rotation parameter $a$. The numerical results are for gravitational quasinormal modes with $l=m=2$. 
The {\it predicted} values (solid line) agree with the {\it numerically} 
computed ones (dashed line) to within $\sim 5 \%$.}
\label{Fig1}
\end{figure}

Figure \ref{Fig1} displays Re$\omega$ for the Kerr black hole, as 
computed numerically in \cite{Ono}, and compare it with the analytically 
conjectured formula Eq. (\ref{Eq4}) \cite{Note2}. We find that the predicted results 
agree with the numerically computed ones to within $\sim 5\%$.

\begin{figure}[tbh]
\centerline{\epsfxsize=9cm \epsfbox{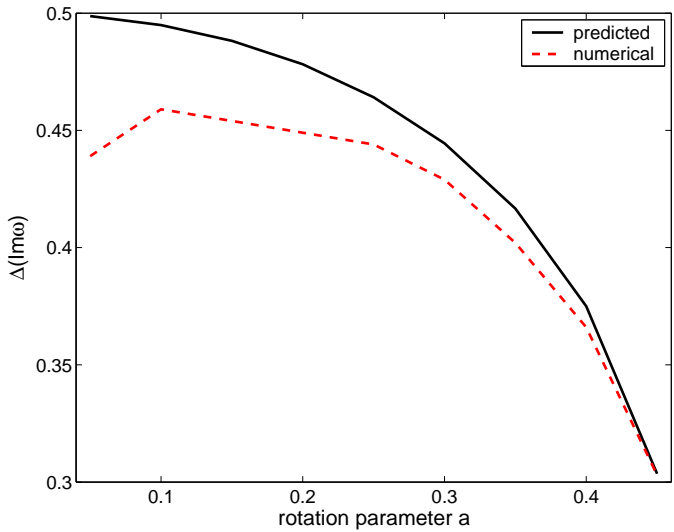}} 
\caption{Spacing of Im$\omega$ as a function of the black hole 
rotation parameter $a$. For $a>0.1$, the {\it predicted} values (solid line) 
agree with the {\it numerically} 
computed ones (dashed line) to within $\sim 7 \%$.}
\label{Fig2}
\end{figure}

Note that the imaginary parts of the asymptotic QNM frequencies are 
{\it equally} spaced in the Schwarzschild case [see Eq. (\ref{Eq1})], 
with a spacing of $1/4M=2\pi T_{BH}$. In order to check if 
this relation holds true for (generic) Kerr black 
holes as well, we display in 
Fig. \ref{Fig2} the spacing $\Delta($Im$\omega)$ using 
the numerical data of \cite{Ono}, and compare it with a 
predicted value of $2\pi T_{BH}$. For $a>0.1$, we find that the 
numerically computed values agree with the predicted ones to within 
$\sim 7\%$.

In summary, based on Bohr's correspondence principle we have conjectured a simple 
formula for the asymptotic QNM frequencies of a generic (rotating) Kerr black hole. We find a 
good agreement between the theoretically predicted frequencies and the numerically computed ones. 
This agreement lands support for the validity of the area spectrum Eq. (\ref{Eq2}).

\bigskip
\noindent
{\bf ACKNOWLEDGMENTS}
\bigskip

I thank a support by the 
Dr. Robert G. Picard fund in physics. 
This research was supported by grant 159/99-3 from the Israel Science Foundation.

\end{document}